\documentclass[fleqn,usenatbib]{mnras}

\usepackage{newtxtext,newtxmath}

\usepackage[T1]{fontenc}
\usepackage{ae,aecompl}
\newcommand{\caps}[1]{{\scshape{#1}}}

\usepackage{graphicx}	
\usepackage{amsmath}	
\usepackage{amssymb}	
\usepackage{color}
\usepackage{placeins}
\usepackage[titletoc]{appendix}

\title[Swift observations of SN 2018cow]{X-ray \textit{Swift} observations of SN 2018cow}

\author[L.E. Rivera Sandoval]{
L.E. Rivera Sandoval$^{1}$\thanks{E-mail: liliana.rivera@ttu.edu},
T.J. Maccarone$^{1}$, A. Corsi$^{1}$, P.J. Brown$^{2}$, D. Pooley$^{3}$, \\
\newauthor J.C. Wheeler$^{4}$ \\
$^{1}$Department of Physics and Astronomy, Box 41051, Science Building, Texas Tech University, Lubbock, TX 79409-1051, USA\\
$^{2}$George P. and Cynthia Woods Mitchell Institute for Fundamental Physics \& Astronomy, Texas A. \& M. \\
University, Department of Physics and Astronomy, 4242 TAMU, College Station, TX 77843, USA\\
$^{3}$Department of Physics and Astronomy, Trinity University, San Antonio, TX, USA ; Eureka Scientific, Inc., USA\\
$^{4}$Department of Astronomy, University of Texas at Austin, Austin, TX, USA\\
}

\pubyear{2018}

\begin{document}
\label{firstpage}
\pagerange{\pageref{firstpage}--\pageref{lastpage}}
\maketitle

\begin{abstract}
Supernova (SN) 2018cow (or AT2018cow) is an optical transient detected in the galaxy CGCG 137--068. It has been classified as a SN due to various characteristics in its optical spectra. The transient is also a bright X-ray source. We present results of the analysis of $\sim 62$\,ks of X-ray observations taken with the \textit{Neil Gehrels Swift Observatory} over 27 d. We found a variable behavior in the $0.3-10$\,keV X-ray light curve of SN\,2018cow, with variability timescales of days. The observed X-ray variability could be due to the interaction between the SN ejecta and a non-uniform circumstellar medium, perhaps related to previous mass ejections from a luminous-blue-variable-like progenitor.  
\end{abstract}




\section{Introduction}
SN\,2018cow (also known as AT2018cow or ATLAS18qqn) is an optical transient detected on 16 June 2018 (10:35:02 UT) with the Asteroid Terrestrial-impact Last Alert System (ATLAS) located at Haleakala and Mauna Loa, Hawaii, USA \citep{11727,2018prentice}. 

The position of the transient (RA=$16^{\rm h}16^{\rm m}00^{\rm s}.22$,  Dec.=+22$^{\circ}$16$'$04$"$.8; J2000) is coincident with that of the galaxy CGCG\,137--068, which is located at a redshift of 0.014\footnote{Value obtained from the NASA/IPAC extragalactic database http://ned.ipac.caltech.edu/forms/byname.html.}, corresponding to a luminosity distance of $59.7\pm 4.2$\,Mpc (assuming $H_0=73.0\pm5$ km s$^{-1}$ Mpc$^{-1}$). SN 2018cow was initially thought to be a cataclysmic variable (CV) star \citep{11727}. However, spectral observations taken on 2018 June 18 UT found a featureless spectrum \citep{11732}. Observations taken on 2018 June 19 UT revealed the presence of Ca II H and K absorption features at the redshift of CGCG\,137--068, thus confirming that SN\,2018cow was in fact an extragalactic transient \citep{11736}. The first X-ray observations of SN\,2018cow taken with the \textit{Neil Gehrels Swift Observatory} also on 2018 June 19 \citep[starting at 10:34:53 UT; ][]{11737,11739} showed that the object had a 0.3--10~keV flux of $(2.6\pm0.3) \times 10^{-11}$ erg cm$^{-2}$ s$^{-1}$, and a hard spectrum with a photon index of $\Gamma=1.6\pm 0.1$. These observations also helped verify the extragalactic nature and to discard the hypothesis that SN\,2018cow was a CV, since at a distance limit of $\sim 700$\,pc derived from its optical luminosity\footnote{Assuming a typical absolute optical magnitude for CVs in outburst $M_V\sim4.5$ \citep{1987-Warner}.} \citep[V=13.8 mags,][]{11737}, the object would have an X-ray luminosity $L_X\sim1.5\times 10^{33}$\,erg\,s$^{-1}$, which is much higher than those of typical CVs in outburst \citep[$L_X\lesssim 10^{32}$\,erg\,s$^{-1}$, see e.g.][]{2001ramsay,2012saito}. Instead, these X-ray observations, suggested that the object could be a Gamma-Ray Burst (GRB) afterglow. GRB afterglows have X-ray luminosities in the range $10^{43}-10^{46}$\,erg\,s$^{-1}$ at $\sim24$\,h after trigger \citep[][]{2012avanzo}, and SN\,2018cow at the distance of $\sim 60$\,Mpc would have an X-ray luminosity $L_X\sim1.15\times 10^{43}$\,erg\,s$^{-1}$. X-ray luminosities of Type Ib/Ic supernovae after few days of the explosion are below $10^{42}$\,erg\,s$^{-1}$ \citep{2016drout}, slightly lower than the observed luminosity of the transient. Alternatively, the X-rays could merely represent a rather extreme shock breakout.

No gamma-rays in the 14--195\,keV range were detected with \textit{Swift}/BAT around the time of optical discovery \citep{11782}. \textit{Fermi}/GBM did not detect activity spatially coincident with SN\,2018cow \citep{11793} in the period 2018 June 13-16, but it is possible that the event occurred outside the field of view of such telescope because the sky region of the transient was observed only 54\% of the time during that period (this could be the case for \textit{Swift}/BAT as well). Additionally, no detection at TeV energies were made in the period 2018 June 13-16\,UT with \textit{HAWC} in the direction of the transient \citep{11792}. 
Ultraviolet observations carried out on 2018 June 19 (10:40:40 UT) with the UVOT instrument on board of \textit{Swift}, showed that SN\,2018cow was also bright at such wavelengths, with Vega magnitude of $11.70\pm0.01$ \citep{11737}.

SN\,2018cow was also identified at millimetre wavelengths with \textit{NOEMA} on 2018 June 20 UT, with a flux of $\sim 6$\,mJy at 90\,GHz \citep{11749}. On 2018 June 22 UT the object had a flux density of $\sim0.5$\,mJy determined using data from \textit{AMI-LA} at a central frequency of $15.5$\,GHz \citep{11774}. \textit{ATCA} detected the transient at 34\,GHz with a flux of $\sim5.6$\,mJy on 2018 June 26 UT \citep{11795}. The flux in that band increased to $\sim7.6$\,mJy two days later \citep{11818}.  

From spectroscopic observations of SN\,2018cow carried out on 2018 June 20 UT with the Xinglong-2.16m, \citet{11740} reported a broad bump or absorption feature (He I [3889] if absorption) and a broad feature at 5040 \AA \ that could be the signature of high velocity blending as in broad-line Type Ic supernovae (SNe BL-Ic). \citet{11753} reported further spectroscopic follow up with \textit{GTC}/ISIRIS on 21 June finding broad undulations similar to SNe BL-Ic, though without a direct match, though absorption blueward of a 5460 \AA \ peak could be FeII at 20,000 km s$^{-1}$. \citet{11766} reported on the evolution of the 4500/5500 \AA \ feature, finding it to have strengthened until 5 d after discovery and then weaking to almost disappear in Liverpool Telescope spectra taken on 24 June.  This would be unusual for an SN emerging from a fading afterglow and/or shock breakout. From spectra taken on 8 July UT, 22 d after discovery, \citet{11836} identified emission features that could be associated to He lines, thus re-classifying SN\,2018cow as a Type Ib SN. 

In this letter we present results of the analysis of 27 d of continuous X-ray follow-up observations of SN\,2018cow with \textit{Swift} in the 0.3--10 keV energy band, starting approximately 3 d after its discovery. 

\section{Observations}

The $\sim 62$\,ks of X-ray \textit{Swift} observations analysed in this letter cover the period from 2018 June 19 to 2018 July 13\,UT. These are a total of 71 observations with exposure times ranging from $\sim 200$\,s to $\sim 3000$\,s. 
The cadence of the X-ray observations is variable, but data were taken at least once per day. We discarded the observation taken on 2018 June 21 at 11:47:14 UT, as the data were heavily affected by a hot column right at the SN\,2018cow position.

All observations were reduced using the \textit{Swift} software\footnote{http://www.swift.ac.uk/analysis/xrt/setup.php}. First, we reprocessed the data using the routine \textsc{xrtpipeline}. Then, the routines \textsc{Xselect} and \textsc{Ximage} were used to measure count rates in the 0.3--10\,keV energy band. We adopted a threshold signal-to-noise ratio of $3$ for detections. To determine rate correction factors due to bad columns we used the routine \textsc{xrtlccorr}. 

For the spectral analysis we followed the \textit{Swift} thread\footnote{http://www.swift.ac.uk/analysis/xrt/spectra.php}. We used \textsc{xrtmkarf} to create the ancillary response files. For the spectral fitting we used the software package \caps{Xspec} \citep[v12.9.1,][]{1996ar}. To model the hydrogen column density N$_H$ we used the model $tbnew\_gas$\footnote{http://pulsar.sternwarte.uni-erlangen.de/wilms/research/tbabs/} with WILM abundances \citep{2000wilm} and VERN cross-sections \citep{1996ver}.
We fitted our spectra with a power-law model (\textsc{pegpwrlw}). The background spectra were extracted from a source free part of the CCD using a circular region with a radius of $195$ arcsec.  For the spectrum presented in Figure \ref{fig:spectrum} the data were grouped to have at least 15 counts per bin.
 
\section{Results}
\label{results}

\begin{figure}
	\includegraphics[width=\columnwidth]{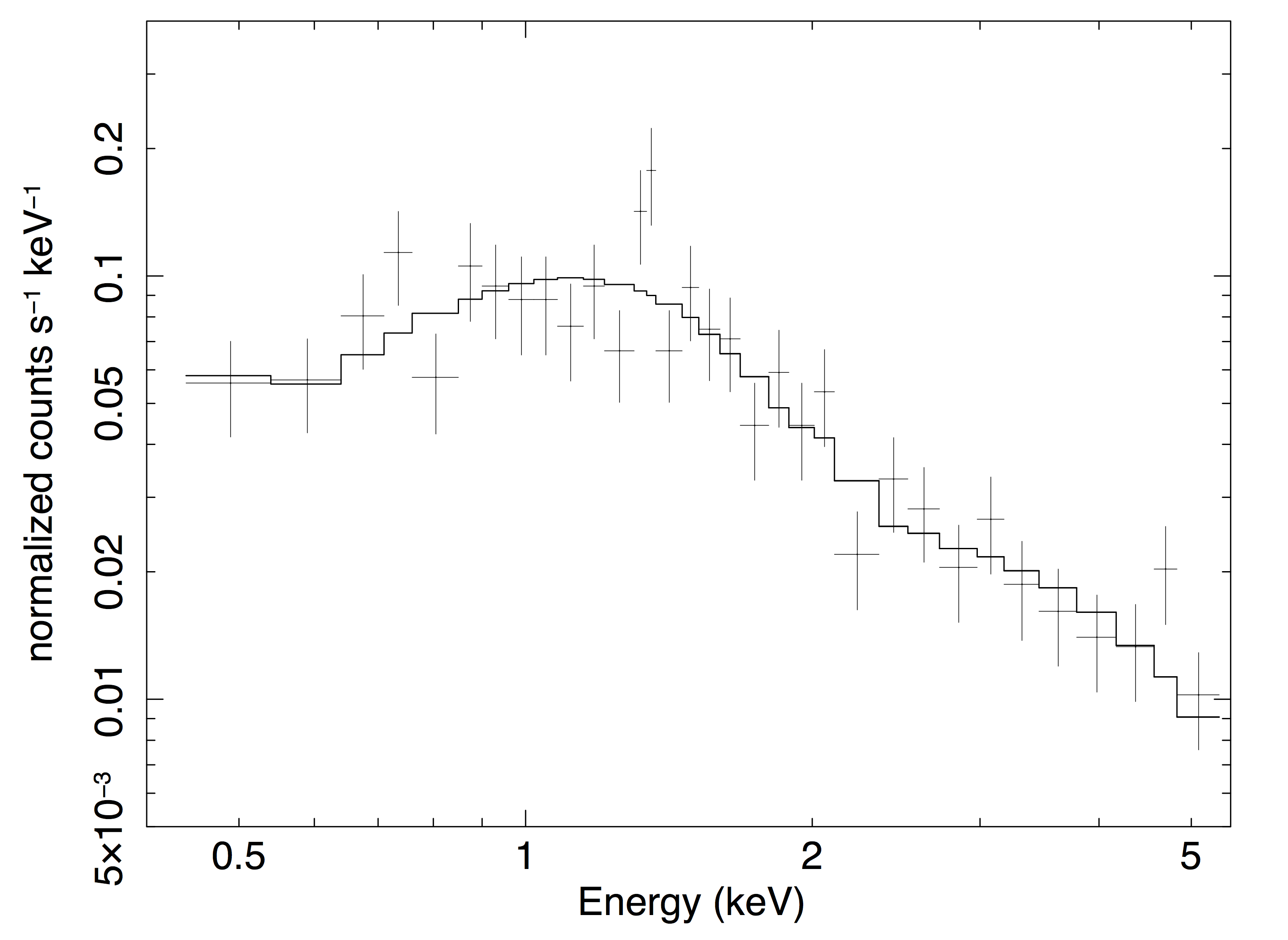}
    \caption{X-ray spectrum of SN 2018cow in the 0.3--10 keV energy band, using an $\sim3$\,ks \textit{Swift} observation taken on 2018 July 7. The spectrum can be described with a photon index $\Gamma=1.4\pm 0.2$. The X-ray flux was $(1.0\pm0.1) \times 10^{-11}$ erg cm$^{-2}$ s$^{-1}$ and the hydrogen column density $N_H=7.6\pm6.1 \times 10^{20}$\,cm$^{-2}$. Quoted errors are at 90\% confidence level.}
    \label{fig:spectrum}
\end{figure}

\begin{figure*}
	\includegraphics[width=\textwidth]{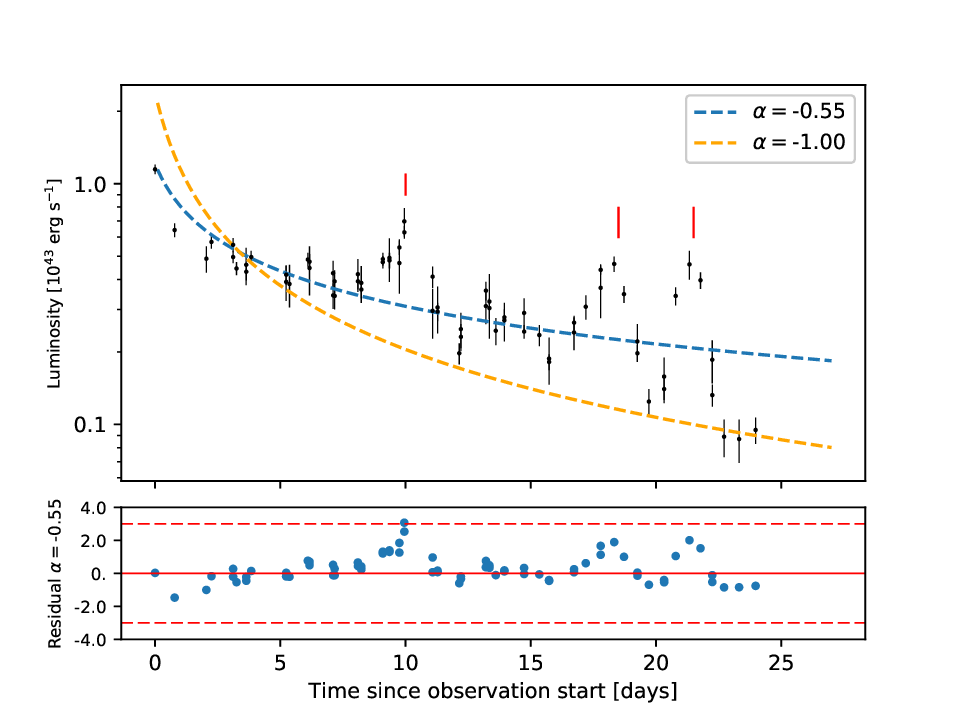}
    \caption{X-ray \textit{Swift} light curve of SN 2018cow from 2018 June 19 (10:34:53 UT) to 2018 July 13 (09:56:36 UT) in the 0.3-10\,keV energy band. A power-law fit with $\alpha=-0.55\pm0.01$ (where $L_X \propto t^{\alpha}$) is shown (blue line). Significant deviations appear during the episodes in which the X-ray unabsorbed luminosity increased. For a comparison, we also plotted a fit with $\alpha=-1$ (orange line). Residuals normalized to the rms error are shown in the bottom part of the plot. The dashed lines indicate the $3\sigma$ level. The presence of consecutive excesses, increases the detection significance of the events (indicated by the red vertical lines). A table with the data is given in the Appendix.}
    \label{fig:lc}
\end{figure*}

\begin{figure}
	\includegraphics[width=\columnwidth, height=\columnwidth]{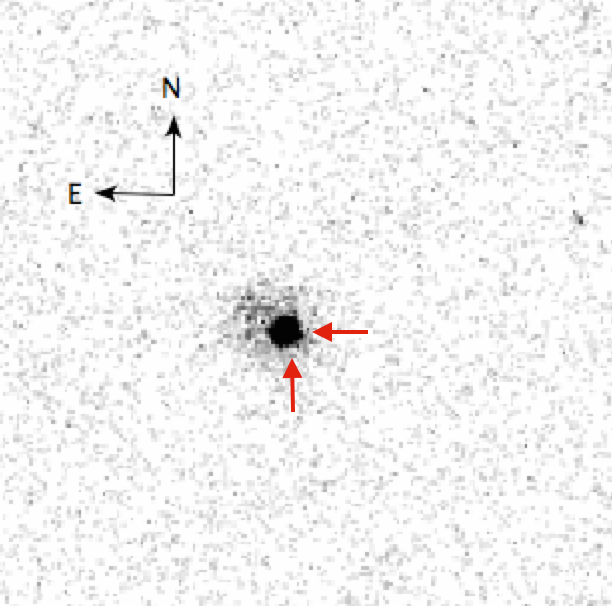}
    \caption{\textit{Swift}/UVOT ultraviolet image of SN 2018cow taken during the X-ray luminosity increase on 2018 June 29 in the filter UVM2. The extended emission surrounding the object is likely due to its host galaxy. The image is $3'\times 3'$. No increase in the UVM2 magnitude was observed during this episode. The arrows indicate the position of SN\,2018cow.}
    \label{fig:finding}
\end{figure}

In Figure \ref{fig:lc} we show the \textit{Swift} 0.3--10~keV light curve of SN\,2018cow from 2018 June 19 UT to 2018 July 13 UT. Data behind this Figure are reported in Table \ref{data_lc}. We see that SN\,2018cow shows a variable behavior on time-scales of days. There are multiple re-brightenings after the first detection on 2018 June 19 UT. The most significant increase in the X-ray count rate is the one detected on 2018 June 29 (09:21:01 UT\footnote{This observation was not affected by a hot column, unlike the longer observation recorded at 09:24:14 UT.}), with a rate of $0.35 \pm0.05$\,counts\,s$^{-1}$(see also Figure \ref{fig:finding}). Fitting an absorbed power-law model to that observation, \citet{11801} determined an X-ray flux in the 0.3--10~keV of $(1.8\pm0.7) \times 10^{-11}$\,erg\,cm$^{-2}$\,s$^{-1}$ and $\Gamma=1.2\pm0.8$ which represented a flux increase of approximately a factor of 2 with respect to the observations taken on 2018 June 21 June (16:37:11 UT), when the object had a flux of $(1.2\pm0.2)\times 10^{-11}$\,erg\,cm$^{-2}$\,s$^{-1}$ and $\Gamma=1.8\pm0.3$ \citep{11761}. 

The X-ray spectrum of SN\,2018cow in the 0.3--10~keV band can be described by a power-law model with $\Gamma\lesssim2$ (Figure \ref{fig:spectrum}) at all epochs. There is no evidence for spectral evolution in the 0.3--10~keV band since discovery, and no evidence for significant spectral evolution during the X-ray increases (measurements are consistent within 90\% errors). The value of the spectral index is in agreement with observations carried out with \textit{Chandra} and \textit{NICER} \citep{11779,11773} and also 
with \textit{NuSTAR} and \textit{INTEGRAL} at higher energies \citep{11775,11813,11843}. Though, indications of a hard component at energies above 15\,keV was found on 2018 June 23 \citep{11775}, which was not longer observed on 2018 July 2 \citep{11813}. The combined spectrum of SN\,2018cow up to 2018 July 13 can be described with a $\Gamma\sim1.6$ and $N_H\sim 7.0\times 10^{20}$\,cm$^{-2}$. 

Assuming a decay of the luminosity $L_X \propto t^{\alpha}$, we found that the temporal evolution of the 0.3--10 keV luminosity can be best fitted with a power-law index $\alpha=-0.55\pm0.01$ (Figure \ref{fig:lc}), which is not consistent with the one found by \textit{INTEGRAL}, where $\alpha=-1$ \citep{11843}. The X-ray increases shown in the light curve of SN\,2018cow seem to significantly deviate from that fit. The presence of consecutive excesses in the residuals, give us confidence on the significance of these events.

\section{Discussion}

SN\,2018cow has been classified as a candidate SN due to its optical spectral properties \citep{11740,11776,11836}. However, an unambiguous classification remains to be established. If indeed SN\,2018cow is a SN, its variable X-ray behaviour (Figure \ref{fig:lc}) could be explained as due to the interaction of the SN shock with non-uniform circum-stellar material (CSM), perhaps related to eruptive, Luminous Blue Variable (LBV)-like mass-loss from the pre-supernova progenitor. Assuming a wind velocity ${\rm v}_{wind}\sim1000$\,km\,s$^{-1}$ \citep[typical for Type Ib/Ic SN; ][]{wellons}, and a SN expansion velocity ${\rm v}_{SN}\sim20,000$\,km\,s$^{-1}$ as determined for SN\,2018cow from optical spectroscopy \citep{11753}, then the $\sim 10$ d between the first two X-ray peaks (probably linked to interactions between the SN and the LBV-like ejecta) would correspond to LBV-like eruptions spaced by $\sim 200$ d. This value is consistent with pulsations of some LBV stars \citep{1998Lamers}. Note, however, that the mass-ejections from the progenitor could have been neither at a constant rate nor at a constant velocity. This would help explain the variable frequency of the observed increases in the X-ray light curve, such as the occurrence of the last one (see Figure \ref{fig:lc}). Considering that the X-ray peaks observed at $\sim10, 18$ and $21$ d after explosion are produced by interactions between the SN and the LBV-like ejecta, the corresponding radii of the shells would be $R={\rm v}_{SN}\times t\sim 1.7 \times 10^{15},\,3.1\times 10^{15}$ and $3.6\times10^{15}$\,cm, respectively\footnote{Assuming a constant velocity. However, \citet{2018prentice} have shown that the velocity of the SN changed significantly within 2 weeks after discovery.}. These values are smaller than those found for Type IIn SNe\footnote{A type of SNe possibly originating from LBV stars.} \citep[$R\sim 5\times 10^{16}$\,cm;][]{2016katsuda} and Type Ib SNe \citep[$R\sim 9\times 10^{15}$\,cm;][]{2008Immler}. 

Fitting an absorbed thermal model (\textsc{bapec}) to the combined spectrum, using solar abundances, a redshift of 0.014 and ${\rm v}_{SN}\sim20,000$\,km\,s$^{-1}$, we obtained a plasma temperature of $10.6^{+2.2}_{-1.2}$\,keV and $N_H=3.6\pm 0.9 \times 10^{20}$\,cm$^{-2}$. We noted that the $N_H$ varies significantly depending on the chosen model. However,  
more observations will allow to better constrain this parameter. Now, assuming that the X-ray luminosity of a thermal plasma is given by the relation $L_X=4\pi R^3(\Delta R/R)\Lambda(T)n^2$ \citep{2008Immler}, where $\Delta R$ is the thickness of the shell ($\Delta R=\Delta t \times {\rm v}_{SN}$, $\Delta t$ is the time interval between the X-ray rise-decline) and $\Lambda(T)$ is the cooling function for an optically thin thermal plasma \citep[$\Lambda(T)\approx 3 \times 10^{-23}$\,erg cm$^3$ s$^{-1}$ for a temperature of $10^8$\,K, consistent with our derived temperature;][]{1976raymond}. Using that approach, we solve for the CSM number density of the shocked plasma inside the emitting shell and obtain $n_{CSM}\sim 3\times10^9$\,cm$^{-3}$. Using this value to estimate the mass that was lost by the progenitor we obtain $>0.08M_\odot$\footnote{That limit is based on the smallest shell. However, a larger mass was likely ejected given the presence of multiple shells.}, which is consistent with LBV mass ejections \citep{2006smith}.

Flux increases in the X-ray light curves of Type Ib/Ic SNe are not commonly observed within few weeks after the explosion. Usually these light curves follow smooth decays \citep[e.g.][]{2009modjaz, 2006campana}. However, there are records of X-ray flux variations in some SNe due to strong interaction with the CSM \citep[e.g.][]{2008Immler}, which supports the idea that the variations observed in the X-ray light curve of SN\,2018cow could be indeed caused by a similar mechanism, though with different factors affecting the timescale of occurrence of the re-brightening episodes. GRB X-ray afterglows also show increases in their light curve. However, they occur at time-scales much shorter than those observed for the SN\,2018cow \citep[see e.g.][]{2006Camp,2018ruffini}. 

In order to determine the nature of SN\,2018cow X-ray emission, we estimated the radio-to-X-ray spectral index comparing the 34\,GHz flux \citep[$\sim5.6$\,mJy; ][]{11795} measured on 2018 June 26 (09:00-14:00 UT), to the \textit{Swift} X-ray flux of two measurements taken the same day within that time interval (13:06:00-13:21:53 UT, $F_X=9.2 \times 10^{-12}$\,erg\,cm$^{-2}$\,s$^{-1}$/$2.4 \times 10^{17}$\,Hz $\approx 4\times 10^{-6}$\,mJy). We obtained $\Gamma_r \approx 1.9$. This value is in agreement with the X-ray spectral index obtained from the \textit{Swift} data. It is also in agreement with results for other CSM-interacting BL-Ic SNe \citep{2014corsi} in which the X-rays seem to have a synchrotron origin. This suggests that a similar mechanism may be powering SN\,2018cow X-ray emission.  

Follow-up observations of SN\,2018cow at all the wavelengths will likely allow to obtain more insights about its nature. However, whatever its classification, SN\,2018cow is already an outstanding object due to the time variability of its X-ray light curve, and to its peculiar multi-wavelength behavior in general. 

\section*{Acknowledgements}
The authors thank the anonymous referee for his/her comments, which improved the manuscript. We also thank the \textit{Swift} team for their fast response to our multiple target of opportunity requests, which provided us with the data used in this paper, and additionally thank Jamie Kennea and Kim Page for further useful advice regarding the \textit{Swift} data. Thanks also to Yuri Cavecchi for comments on the paper.




\bibliographystyle{mnras}
\bibliography{biblio} 



\clearpage
\appendix
\label{appendix}
\section{Swift X-ray light curve data}
We present the \textit{Swift} data used to generate Figure \ref{fig:lc} in Section \ref{results}. 

\begin{table}
\centering
\caption{\textit{Swift} X-ray light curve of SN\,2018cow in the 0.3--10 keV. The time since the first observation (2018 June 19 at 10:34:53 UT) is indicated in column 1. We have used a factor of $1.97\times10^{43}$ erg s$^{-1}$
to convert the count rates to unabsorbed luminosities.}
\label{data_lc}
\begin{tabular}{c c c}
t & Luminosity & $\sigma_{L_X}$ \\
(d) & ($10^{43}$ erg s$^{-1}$) & ($10^{43}$ erg s$^{-1}$)\\
\hline
\hline
0.000	&	1.147	&	0.051	\\
0.781	&	0.642	&	0.041	\\
2.047	&	0.488	&	0.061	\\
2.252	&	0.573	&	0.036	\\
3.116	&	0.557	&	0.038	\\
3.116	&	0.496	&	0.028	\\
3.251	&	0.444	&	0.028	\\
3.639	&	0.460	&	0.081	\\
3.641	&	0.431	&	0.034	\\
3.838	&	0.496	&	0.030	\\
5.233	&	0.391	&	0.065	\\
5.236	&	0.419	&	0.038	\\
5.366	&	0.383	&	0.077	\\
6.099	&	0.484	&	0.032	\\
6.165	&	0.446	&	0.103	\\
6.168	&	0.474	&	0.036	\\
7.105	&	0.425	&	0.053	\\
7.110	&	0.344	&	0.041	\\
7.167	&	0.393	&	0.043	\\
7.170	&	0.342	&	0.041	\\
8.096	&	0.421	&	0.065	\\
8.099	&	0.393	&	0.030	\\
8.231	&	0.387	&	0.067	\\
8.234	&	0.363	&	0.030	\\
9.091	&	0.472	&	0.028	\\
9.093	&	0.486	&	0.030	\\
9.356	&	0.492	&	0.101	\\
9.358	&	0.480	&	0.036	\\
9.752	&	0.468	&	0.118	\\
9.754	&	0.543	&	0.038	\\
9.949	&	0.697	&	0.095	\\
9.951	&	0.628	&	0.038	\\
\end{tabular}
\end{table}

\FloatBarrier

\addtocounter{table}{-1}
\begin{table}
\centering
\caption{Continued}
\begin{tabular}{c c c}
t & Luminosity & $\sigma_{L_X}$ \\
(d) & ($10^{43}$ erg s$^{-1}$) & ($10^{43}$ erg s$^{-1}$)\\
\hline
\hline
11.078	&	0.296	&	0.069	\\
11.080	&	0.411	&	0.041	\\
11.277	&	0.306	&	0.067	\\
11.279	&	0.294	&	0.026	\\
12.143	&	0.197	&	0.020	\\
12.207	&	0.249	&	0.041	\\
12.209	&	0.231	&	0.030	\\
13.205	&	0.310	&	0.049	\\
13.209	&	0.359	&	0.032	\\
13.346	&	0.324	&	0.097	\\
13.348	&	0.304	&	0.030	\\
13.601	&	0.245	&	0.032	\\
13.941	&	0.271	&	0.049	\\
13.945	&	0.278	&	0.022	\\
14.731	&	0.290	&	0.043	\\
14.734	&	0.243	&	0.016	\\
15.339	&	0.235	&	0.024	\\
15.728	&	0.188	&	0.041	\\
15.730	&	0.182	&	0.014	\\
16.724	&	0.241	&	0.038	\\
16.727	&	0.265	&	0.018	\\
17.194	&	0.308	&	0.036	\\
17.788	&	0.369	&	0.093	\\
17.789	&	0.438	&	0.020	\\
18.329	&	0.464	&	0.034	\\
18.719	&	0.348	&	0.028	\\
19.253	&	0.221	&	0.039	\\
19.255	&	0.197	&	0.016	\\
19.714	&	0.124	&	0.016	\\
20.320	&	0.158	&	0.032	\\
20.322	&	0.140	&	0.018	\\
20.783	&	0.342	&	0.030	\\
21.333	&	0.462	&	0.063	\\
21.773	&	0.397	&	0.032	\\
22.242	&	0.186	&	0.038	\\
22.244	&	0.132	&	0.014	\\
22.715	&	0.089	&	0.016	\\
23.315	&	0.087	&	0.018	\\
23.973	&	0.095	&	0.012	\\
\end{tabular}
\end{table}


\bsp	
\label{lastpage}
\end{document}